\documentclass[5p]{elsarticle}

 \usepackage{graphicx}
 \usepackage[font=footnotesize]{subfig} 
 \usepackage{relsize}
 \usepackage{subfig}

\usepackage{amssymb}
\usepackage[pdftex,dvipsnames,usenames]{color}		

\def\beq{\begin{equation}}
\def\eeq{\end{equation}}
\def\bea{\begin{eqnarray}}
\def\eea{\end{eqnarray}}

\def\bigD{{\cal D}}

\newcommand*{\eqref}[1]{Eq.~(\ref{eq:#1})}
\newcommand*{\eqlab}[1]{\label{eq:#1}}
\newcommand*{\figref}[1]{Fig.~\ref{fig:#1}}
\newcommand*{\figlab}[1]{\label{fig:#1}}

\def\VYP#1#2#3{{\bf #1}, #3 (#2)}  

\newcommand{\Omit}[1]{}

\begin{document}

\begin{frontmatter}



\title{The cosmic-ray air-shower signal in Askaryan radio detectors}

\author[1]{Krijn D. de Vries}
\ead{krijndevries@gmail.com}
\author[1]{Stijn Buitink}
\author[1]{Nick van Eijndhoven}
\author[2]{Thomas Meures}
\author[2]{Aongus \'O Murchadha}
\author[1,3]{Olaf Scholten}
\address[1]{Vrije Universiteit Brussel, Dienst ELEM, B-1050 Brussels, Belgium}
\address[2]{Universit\'e Libre de Bruxelles, Department of Physics, B-1050 Brussels, Belgium}
\address[3]{University Groningen, KVI Center for Advanced Radiation Technology,Groningen, The Netherlands}


\begin{abstract}
We discuss the radio emission from high-energy cosmic-ray induced air showers hitting Earth's surface before the cascade has died out in the atmosphere. The induced emission gives rise to a radio signal which should be detectable in the currently operating Askaryan radio detectors built to search for the GZK neutrino flux in ice. The in-air emission, the in-ice emission, as well as a new component, the coherent transition radiation when the particle bunch crosses the air-ice boundary, are included in the calculations.
\end{abstract}

\end{frontmatter}
\section{Introduction}
We calculate the radio emission from cosmic-ray-induced air showers as a possible (background) signal for the Askaryan radio-detection experiments currently operating at Antarctica~\cite{ANITA,ARA,ARIANNA}. A high-energy neutrino interacting in a medium like (moon)-rock, ice, or air will induce a high-energy particle cascade. In 1962 Askaryan predicted that during the development of such a cascade a net negative charge excess arises mainly due to Compton scattering~\cite{Ask62}. This net excess charge by itself will induce a radio signal that can be used to measure the original neutrino. This Askaryan radio emission~\cite{Ask62,Zas92,Mun97} has been confirmed experimentally at SLAC~\cite{Sal01}, and more recently the Askaryan effect was also confirmed in nature by the radio emission from cosmic-ray induced air showers~\cite{Mar11,Aab14,Sch15}.

For high-energy cosmic-ray air showers, along with the Askaryan emission, there is another emission mechanism due to a net transverse current that is induced in the shower front by Earth's magnetic field [11-14].
Recently the radio emission from cosmic-ray air showers has been measured in great detail by the LOFAR collaboration~\cite{Sch15,Nel15,Bui15}, confirming the predictions from several independent radio emission models [17-20].

Most Askaryan radio detectors [1-3,21-23] 
search for so-called GZK neutrinos that are expected from the interaction of ultra-high-energy cosmic-ray protons with the cosmic microwave background~\cite{Grei66,Zat66}. The expected GZK neutrinos are extremely energetic with energies in the EeV range, while the flux at these energies is expected to fall below one neutrino interaction per cubic kilometer of ice per year. Therefore, to detect these neutrinos an extremely large detection volume, even larger than the cubic kilometer currently covered by the IceCube experiment, is needed. Due to its long attenuation length, the induced radio signal is an excellent means to detect these GZK neutrinos. This has led to the development of several radio-detection experiments~[1-6,26-30]. Nevertheless, the highest-energy neutrinos detected so-far are those observed recently by the IceCube collaboration~\cite{I3_2013sc} and have energies up to several PeV, just below the energies expected from the GZK neutrino flux.

In this article we calculate the radio emission from cosmic-ray-induced air showers as a possible (background) signal for the Askaryan radio-detection experiments currently operating at Antarctica~\cite{ANITA,ARA,ARIANNA}. Besides the emission during the cascade development also transition radiation should be expected when the cosmic ray air shower hits Earth's surface~\cite{Mar86,Mar14}. It follows that the induced emission is very hard to distinguish from the direct Askaryan emission from a high-energy neutrino induced cascade in a dense medium such as ice.

\section{Radio emission from a particle cascade}
We start from the Li\'enard-Wiechert potentials for a point-like four current from classical electrodynamics and closely follow the macroscopic MGMR~\cite{dVries10} and EVA~\cite{Wer12} models. Both models were developed to describe the radio emission from cosmic-ray-induced air showers. The Li\'enard-Wiechert potentials for a point charge, $A^{\mu}_{PL}(t,\vec{x})$, as seen by an observer positioned at $\vec{x}$ at an observer time $t$ are obtained directly from Maxwell's equations after fixing the Lorenz gauge~\cite{Jackson},
\beq
A^{\mu}_{PL}(t,\vec{x})=\left. \frac{1}{4\pi\epsilon_0}\frac{J^{\mu}}{|\bigD|}\right|_{ret} \;.
\eeq
The point-like current is defined by $J^{\mu}=e V^{\mu}$, where $e$ is the charge, and $V^{\mu}$ is the four-velocity for a particle at $\vec{\xi}(t_r)$ where the retarded emission time is denoted by $t_r$. The denominator of the vector potential, $\bigD$, is the retarded four-distance. For an extended current with longitudinal dimension $h$ and lateral dimensions $\vec{r}$, the vector potential has to be convolved with the charge distribution given by the weight function $w(\vec{r},h)$,
\beq
A^{\mu}(t,\vec{x})=\frac{1}{4\pi\epsilon_0}\int \;\mathrm{d}h\;\mathrm{d^2}r\; \left. \frac{J^{\mu}w(\vec{r},h)}{|\bigD|} \right|_{ret},
\eeq
where the vector potential has to be evaluated at the retarded emission time $t_r$.
\begin{figure}[!ht]
\centerline{
\includegraphics[width=.5\textwidth, keepaspectratio]{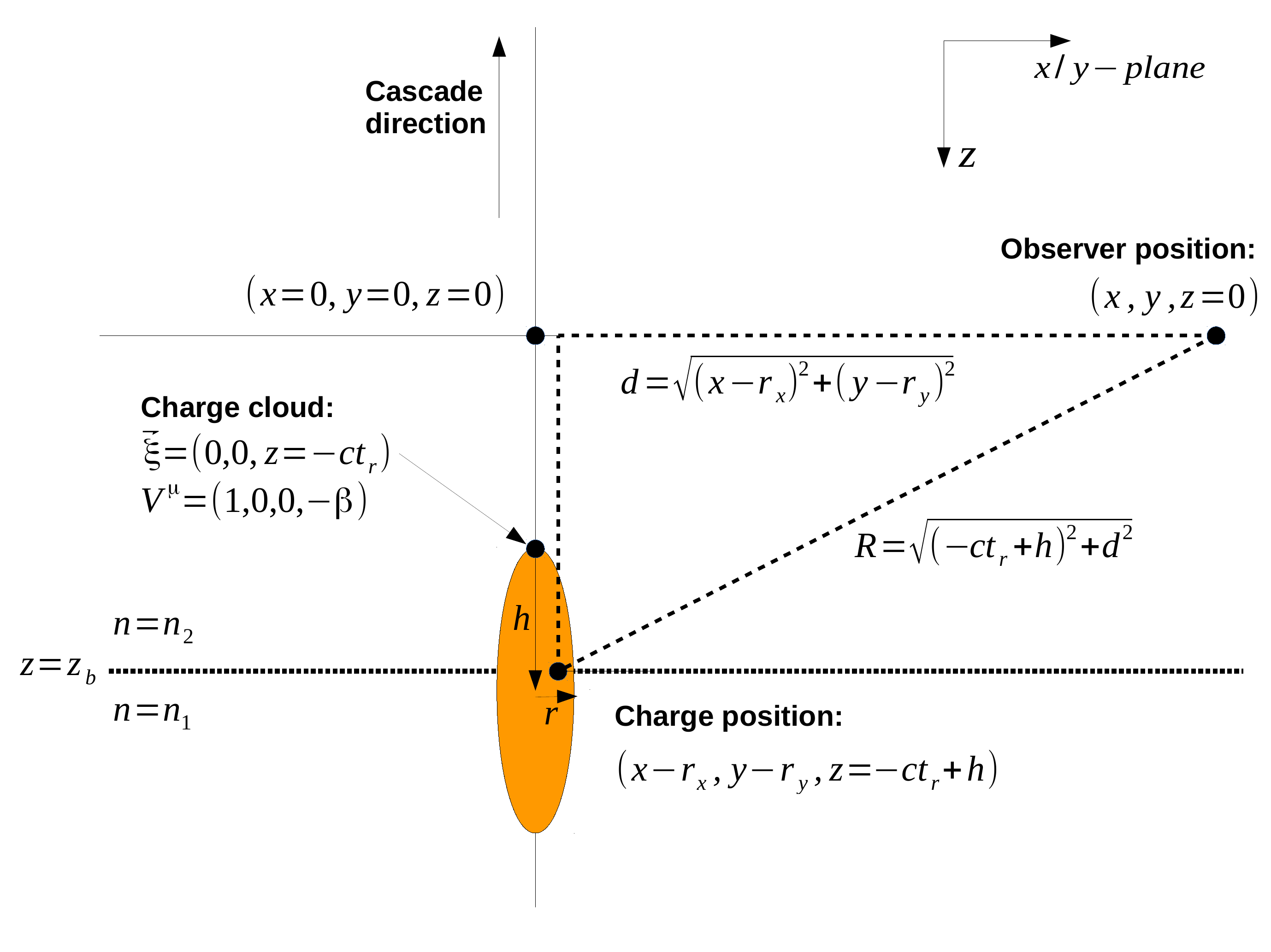}}
\caption{The geometry used to calculate the radiation emitted from a charge cloud crossing a boundary at $z=z_b$. The observer is positioned at an impact parameter $d=\sqrt{(x-r_x)^2+(y-r_y)^2}$.}
\figlab{geometry}
\end{figure}
The corresponding geometry is denoted in~\figref{geometry}. We consider an observer positioned at an impact parameter $d=\sqrt{(x-r_x)^2+(y-r_y)^2}$ perpendicular to the charge track, where $r_x$, and $r_y$ denote the lateral position of the considered charge within the charge cloud. Defining the element in the plane of the observer perpendicular to the charge-track as $z=0$, we can define the time at which the front of the charge cloud crosses this plane to be $t=0$. Using these definitions the position of the charge along the track is now given by $z=-ct_r+h$.

Fixing the geometry, the vector potential can now be evaluated. The retarded emission time is obtained from the light-cone condition with respect to the optical path length $L$,
\beq
c(t-t_r)=L \;,
\eqlab{ctr}
\eeq
from which the relation between the observer time and the emission time, $t_r(t)$, can be obtained. It should be noted that $t_r$ is a negative quantity. For a medium consisting out of $m$ layers with different index of refraction $n_i$, the optical path length can be defined by
\beq
L=\sum_{i=1}^m n_i d_i,
\eeq
where the distance $d_i$, the distance covered by the emission in layer $i$, is obtained by using a ray-tracing procedure based on Snell's law. Following~\cite{Wer08}, the retarded distance for a signal traveling through different media is given by,
\beq
\bigD=L\frac{\mathrm{d}t}{\mathrm{d}t_r} \;.
\eqlab{rv1}
\eeq
In this work the index of refraction is assumed to be independent of frequency within the radio frequency range starting from a few MHz, up to several GHz. In the simplified situation where the signal travels through a medium with constant index of refraction $n$, the retarded distance can be written in the more common form,
\beq
\bigD=nR(1-n\beta\cos(\theta)) \;,
\eqlab{rv2}
\eeq
where $\theta$ denotes the opening angle between the line of sight from the emission point to the observer and the direction of movement of the emitting charge.

\subsection{Cherenkov effects for a single electron}
For a single electron moving at a highly relativistic velocity $\vec{\beta}=\vec{v}/c\approx1$ along the $z$-axis (by definition), the current is given by $J^{\mu}=e\,(1,0,0,-\beta)$. The electric field is now obtained directly from the Li\'enard-Wiechert potentials through,
\beq
E^i(t,\vec{x})=-\frac{\mathrm{d}A^{0}}{\mathrm{d}x^i}-\frac{\mathrm{d}A^i}{\mathrm{d}ct},
\eeq
where $i=x,y$ gives the polarization of the field in the transverse direction, and $x^{i}$ denotes the observer position in the transverse plane ($x^1=x, x^2=y$). For the moment we will ignore the electric field in the longitudinal direction and, since $A^i\propto J^i=0$ for $i=1,2$ (there is no transverse current), we only have to consider the spatial derivative of the scalar potential. The electric field in the longitudinal direction will in general be small and can easily be calculated following the gauge condition $\vec{k}\cdot\vec{\epsilon}=0$, where $\vec{k}$ is the momentum vector of the photon and $\vec{\epsilon}$ the polarization. Hence the photon cannot be polarized along its direction of motion. Starting at the zeroth component of the vector potential, the spatial derivative can be evaluated by,
\beq
\frac{\mathrm{d}A^{0}}{\mathrm{d}x^i}=\frac{\mathrm{\partial}}{\mathrm{\partial} x^i} A^0,
\eqlab{Efield}
\eeq
which corresponds to the radiation from a net charge moving through the medium. For a relativistic electron ($\beta\approx 1$) moving in a medium with a refractive index $n>1$ this term becomes,
\bea
E_{st}^i(t,\vec{x})&&=-\frac{\mathrm{\partial}}{\mathrm{\partial} x^{i}}A^{0}\nonumber\\
&&=\frac{-e}{4\pi\epsilon_0}\frac{(1-n^2) x^i}{|\bigD|^3}.
\eea
Where the label 'st', denotes that the field is due to a highly relativistic non time-varying steady charge. The emission shows a radial polarization direction and vanishes linearly with the distance of the observer to the shower core. This component of the electric field is suppressed by the factor $1-n^2$, which vanishes in vacuum. In a medium with an index of refraction larger than unity however, this factor does not vanish and Cherenkov radiation is observed at the point where the retarded distance vanishes, $\bigD=\sqrt{t^2+(1-n^2\beta^2)(x^2+y^2)}=0$. 

The retarded distance vanishes at the finite Cherenkov angle $\cos(\theta_{CH})=\frac{1}{n\beta}$ (see \eqref{rv2}) where the electric field diverges. One intuitive way to understand the Cherenkov effect follows from the more general definition of $\bigD$ given in~\eqref{rv1}. For a vanishing retarded distance, the derivative $\mathrm{d}t/\mathrm{d}t_r$ has to vanish. It follows that the function $t(t_r)$ is flat at this point. Hence at an observer time $t$, signals emitted at different emission times $t_r$ will be observed at once, leading to a boosted electric field. The vanishing of the retarded distance leads to a divergence in the electric field expressions. These divergences are integrable and therefore disappear for coherent emission by performing an integration over the finite charge and current distributions in the shower front~\cite{Wer12}.

\subsection{Transition radiation for a single electron}
So far we calculated the component of the electric field due to a relativistically moving net charge in a medium with a refractive index equal to $n$. How does this compare to the transition radiation for a relativistic charge crossing from a medium with refractive index $n_1$ to a medium with refractive index $n_2$? The vector potential for a single electron now becomes,
\bea
A^{0}(t,\vec{x})=&&\frac{e}{4\pi\epsilon_0}\left(\frac{x^i}{|\bigD|}\theta(z-z_b)\right.\nonumber\\
&&\;\;\;\;\;\;\;\;\left.+\frac{x^i}{|\bigD|}\theta(z_b-z)\right)\;,
\eea
where the discontinuity at a distance $z_b=-ct_b$, corresponding to the retarded emission time $t_b$ when the electron crosses the boundary, is reflected by the step function $\theta(x)$ which is defined by,
\beq
\theta(x)=\left\{ \begin{array}{ll}
         0 & \mbox{if $x < 0$}\\
         1 & \mbox{if $x > 0$} \end{array} \right. \;.
\eeq
Since the step function is a function of the retarded emission time,
\beq
\theta(z-z_b)=\theta(-c (t_r-t_b)) \;,
\eeq
an additional term has to be added to~\eqref{Efield}. The full electric field is now given by,
\beq
\frac{\mathrm{d}A^{0}}{\mathrm{d}x^i}=\frac{\mathrm{\partial}}{\mathrm{\partial} x^i} A^0 +\frac{\mathrm{\partial} t_r}{\mathrm{\partial} x^i}\frac{\mathrm{\partial}}{\mathrm{\partial} t_r} A^0,
\eqlab{etr}
\eeq
where in case of a single electron the second term on the right hand side of~\eqref{etr} will correspond to the transition radiation. The transition radiation can therefore be evaluated as,
\bea
&&E_{tr}^i(t,\vec{x})=\frac{\mathrm{\partial} t_r}{\mathrm{\partial} x^i} \frac{\mathrm{\partial}}{\mathrm{\partial} t_r} A^0\nonumber\\
&&=\frac{e\delta(c (t_r-t_b))}{4\pi\epsilon_0 c}\lim_{\epsilon\rightarrow 0}\left( \frac{x^i}{|\bigD|^2_{t_r+\epsilon}} -\frac{x^i}{|\bigD|^2_{t_r-\epsilon}} \right).
\eqlab{etrr}
\eea
It follows that when there is no boundary, hence $n_1=n_2$, the transition radiation vanishes as it should. Looking more closely at the obtained expression in~\eqref{etrr}, it can be described as the superposition of the emission just before the particle crosses the boundary and the field just after the particle crossed the boundary. The two terms interfere destructively. This corresponds well to the mirror-charge approach for determining the transition radiation as applied by Ginzburg et al.~\cite{Gin90} and the expressions obtained in~\cite{Veen10,Jam11} and references therein.

\begin{figure}[!ht]
\centerline{
\includegraphics[width=.5\textwidth, keepaspectratio]{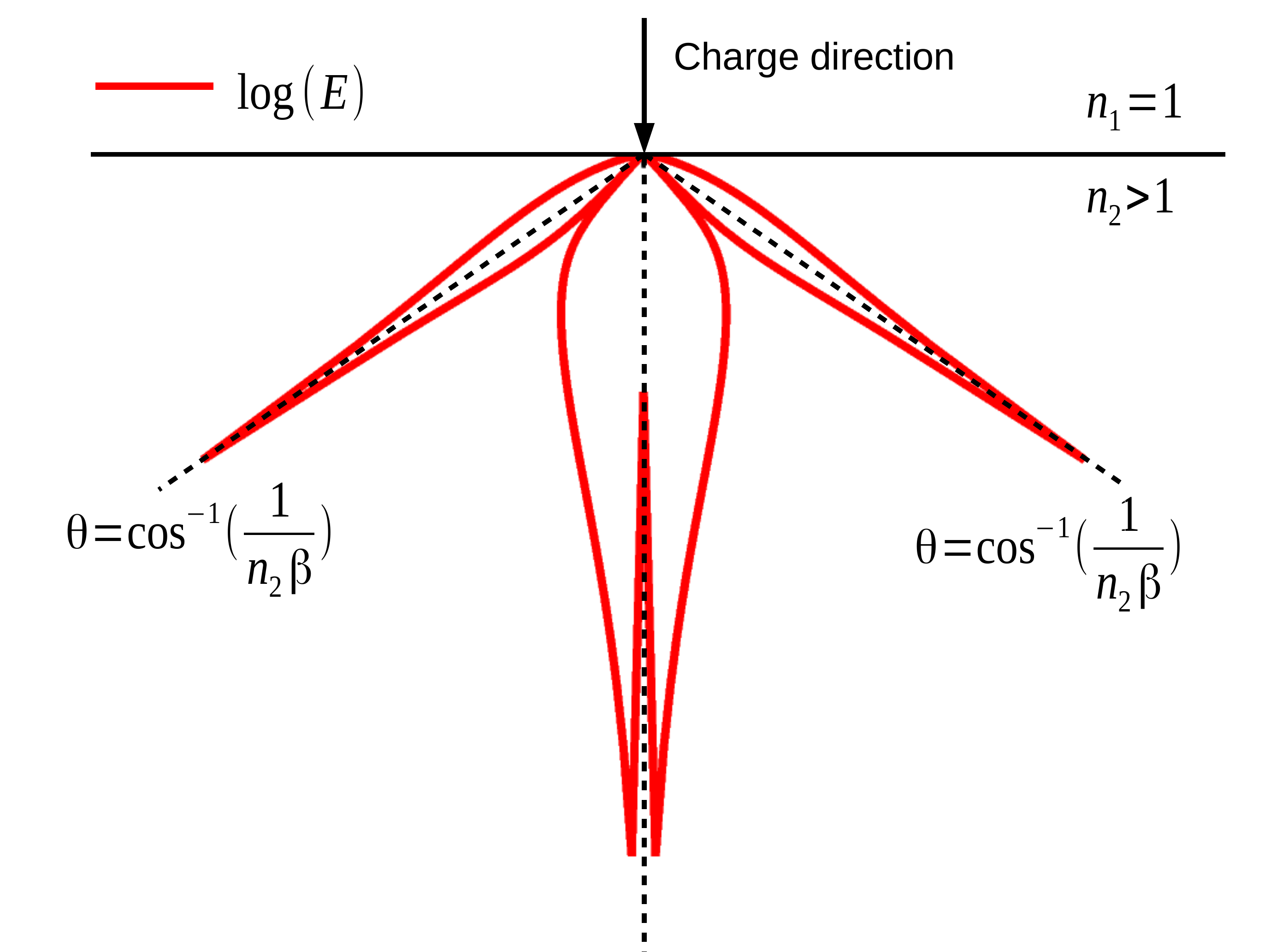}}
\caption{The angular intensity of the transition radiation seen from a charge moving from vacuum, $n_1=1$, to a dense medium $n_2 > 1$.}
\figlab{tr_angle}
\end{figure}

In~\figref{tr_angle} the intensity of the transition radiation is shown as function of angle. Since in vacuum the retarded distance vanishes at forward angles the first term dominates in the forward direction. At very small angles the intensity is suppressed due to the factor {x} in \eqref{etrr}. With increasing angle $1/ |\bigD|^2_{t_r+\epsilon}$ drops in magnitude while the contribution from below the surface, proportional to $1/ |\bigD|^2_{t_r-\epsilon}$ increases in magnitude. Since the two interfere destructively there is a cancellation at a certain angle after which the second term dominates. At the Cherenkov angle in the medium, $\theta_2$, the second term diverges. 
\subsubsection{The sudden appearance signal}
Another effect which is similar to transition radiation is the sudden appearance signal. One example of such a signal is found in accelerator experiments~\cite{Shi94,Gor20}. Here the charged particle beam is accelerated, but during the acceleration process the charge is (partly) blocked from an observer. The consequence is that when the beam leaves the accelerator, the observer suddenly observes a charge. This effect can be described in a similar way as transition radiation. The vector potential simply becomes,
\bea
A^{0}(t,\vec{x})=&&\frac{e}{4\pi\epsilon_0}\frac{1}{|\bigD|}\theta(z_b-z),
\eea
for a beam traveling in a medium with refractive index $n$. Following the transition radiation calculation, the electric field is now directly obtained by,
\bea
E_{sa}^i(t,\vec{x})&&=\frac{\mathrm{\partial} t_r}{\mathrm{\partial} x^i} \frac{\mathrm{\partial}}{\mathrm{\partial} t_r} A^0\nonumber\\
&&=\lim_{\epsilon\rightarrow 0}\frac{e\delta(c (t_r-t_b))}{4\pi\epsilon_0 c}\frac{x^i}{|\bigD|^2_{t_r+\epsilon}}.
\eea
It should be noted that the delta-function is a function of the retarded emission time, $t_r$. The functional dependence can be shifted to the observer time, $t$, after which the field is given by the more common expression,
\beq
E^i_{sa}(t,\vec{x})=\frac{e \delta(ct+z_b-L_b) }{4\pi\epsilon_0c}\frac{x^i}{LD},
\eeq
where $L_b$ denotes the optical path length for the signal emitted at the boundary point toward the observer.

\subsection{Time varying current emission}
So far we considered radiation from a single electron. In case of an electron bunch, there will be another radiation component due to the time variation of the total number of charges. In general this time variation can be linked to the net contribution of coherent bremsstrahlung emission of charges dropping out of the high-energy charge cloud and the emission of Compton electrons which are suddenly accelerated to relativistic speed. More generally, we can define the total number of particles at the retarded emission time $t_r$ by the distribution $N_e(t_r)$. Defining the four-current as,
\beq
J^{\mu}(t_r)=e N_e(t_r)V^{\mu} \;,
\eeq
the vector potential becomes,
\beq
A^{\mu}(t,\vec{x})=\left. \frac{1}{4\pi\epsilon_0}\frac{J^{\mu}(t_r)}{|\bigD|}\right|_{ret},
\eeq
which gives the point-like vector potential for a non extended current. For a cosmic-ray air shower, the two main emission mechanisms are due to a time-varying transverse current which is induced by Earth's magnetic field, and the Askaryan emission due to the time-variation of the net negative charge-excess in the cascade. Nevertheless, for the geometry considered in this article, describing the emission for a perpendicular incoming shower hitting the ice surface at the South-Pole, the cascade will be aligned with Earth's magnetic field and the transverse current vanishes. Therefore, in this section we focus on the emission from a time varying charge. For more information about the radio emission arising from the transverse current the reader is referred to~\cite{Wer12,dVries10}.

In case of a point-like time varying charge $Q(t_r)=eN_e(t_r)$, the partial derivative with respect to the retarded time of the second term on the right hand side of~\eqref{etr} will now also get a contribution where the derivative acts on this current. This leads to the varying charge emission,
\beq
E_{vc}^i(t,\vec{x})=\frac{-1}{4\pi\epsilon_0c}\frac{n x^i}{|\bigD|^2}\frac{dQ}{dt_r},
\eeq
which is the far-field radiation component. It should be noted that also here the signal scales like $1/|\bigD|^2$ and hence Cherenkov or equivalently relativistic time-compression effects apply equally well for this component of the radiation. Furthermore, a similar polarization behavior as for the emission from a highly relativistic steady charge as well as the transition radiation is obtained.

\section{Coherent emission}
To obtain the coherent emission we need to consider the spatial extent of the particle cascade. This is done by inclusion of the weight function $w(\vec{r},h)$. The weight function is normalized such that $\int \mathrm{d}h\;\mathrm{d^2}r\, w(\vec{r},h)=1$. In a realistic situation, there will also be emission from the charged trail which is left behind after the cascade has passed. A detailed calculation including this can be found in~\cite{dVries10}. The expressions given below for the coherent emission are obtained including this positive trail.

The electric field is now obtained by convolving the point-like current with the particle distributions in the shower front which can be evaluated numerically,
\bea
\vec{E}_{st}(t,\vec{x})&=&\frac{-e\,d}{4\pi\epsilon_0}\int \mathrm{d}h\;\mathrm{d^2}r \frac{(1-n^2)}{|\bigD|^3}\nonumber\\
&\times&N_e(t_r) w(\vec{r},h)\,\hat{p}\\
\vec{E}_{vc}(t,\vec{x})&=&\frac{-e\,d}{4\pi\epsilon_0 c}\int \mathrm{d}h\;\mathrm{d^2}r \frac{n}{|\bigD|^2}\nonumber\\
&\times&w(\vec{r},h)\frac{\mathrm{d}N_e(t_r)}{\mathrm{d}t_r}\,\hat{p} \;.
\eea
Here $\hat{p}=\vec{e}_r \times( \vec{e}_r \times \vec{e}_\beta)$ is the polarization of the signal, $\vec{e}_r$ is the unit vector pointing from the emission point to the observer, and $\vec{e}_\beta$ is the unit vector denoting the direction of the cascade. For the transition radiation the delta-function in~\eqref{etrr} can be rewritten as,
\beq
\delta(z-z_b)=\delta(h-c(t_r-t_b))\;.
\eeq
The electric field can now be solved analytically by integrating the delta-function and is given by,
\bea
\vec{E}_{tr}(t,\vec{x})&=&
\lim_{\epsilon\rightarrow 0}\int \mathrm{d}h\;\mathrm{d^2}r \left[ \frac{e\, d\, N_e(t_r)\,w(\vec{r},h)}{4\pi\epsilon_0 c}\right.\nonumber\\
&\times&\left.\left( \frac{1}{|\bigD|^2_{t_r-\epsilon}}-\frac{1}{|\bigD|^2_{t_r+\epsilon}}\right)\right] \nonumber\\
&\times&\delta(h-c(t_r-t_b))\,\hat{p}\nonumber\\
&=&\lim_{\epsilon\rightarrow 0}\int \mathrm{d^2}r\frac{e\,d\,N_e(t_r)\,w(\vec{r},h)}{4\pi\epsilon_0 c}\nonumber\\
&\times&\left.\left( \frac{1}{|\bigD|^2_{t_r-\epsilon}}-\frac{1}{|\bigD|^2_{t_r+\epsilon}}\right)\hat{p}\right|_{h=c(t_r-t_b)}\;.
\eea
Following the same procedure the sudden appearance signal is given by,
\beq
\vec{E}_{sa}(t,\vec{x}) = \lim_{\epsilon\rightarrow 0}\int \mathrm{d^2}r \left. \frac{e\,d\,N_e(t_r)\,w(\vec{r},h)}{4\pi\epsilon_0 c\,|\bigD|^2_{t_r+\epsilon}}\hat{p}\right|_{h=c(t_r-t_b)}
\eeq

\section{The cosmic-ray air shower signal in Askaryan radio detectors}

\begin{figure}[ht!]
  \subfloat[]{\figlab{profilea}\includegraphics[width=.5\textwidth]{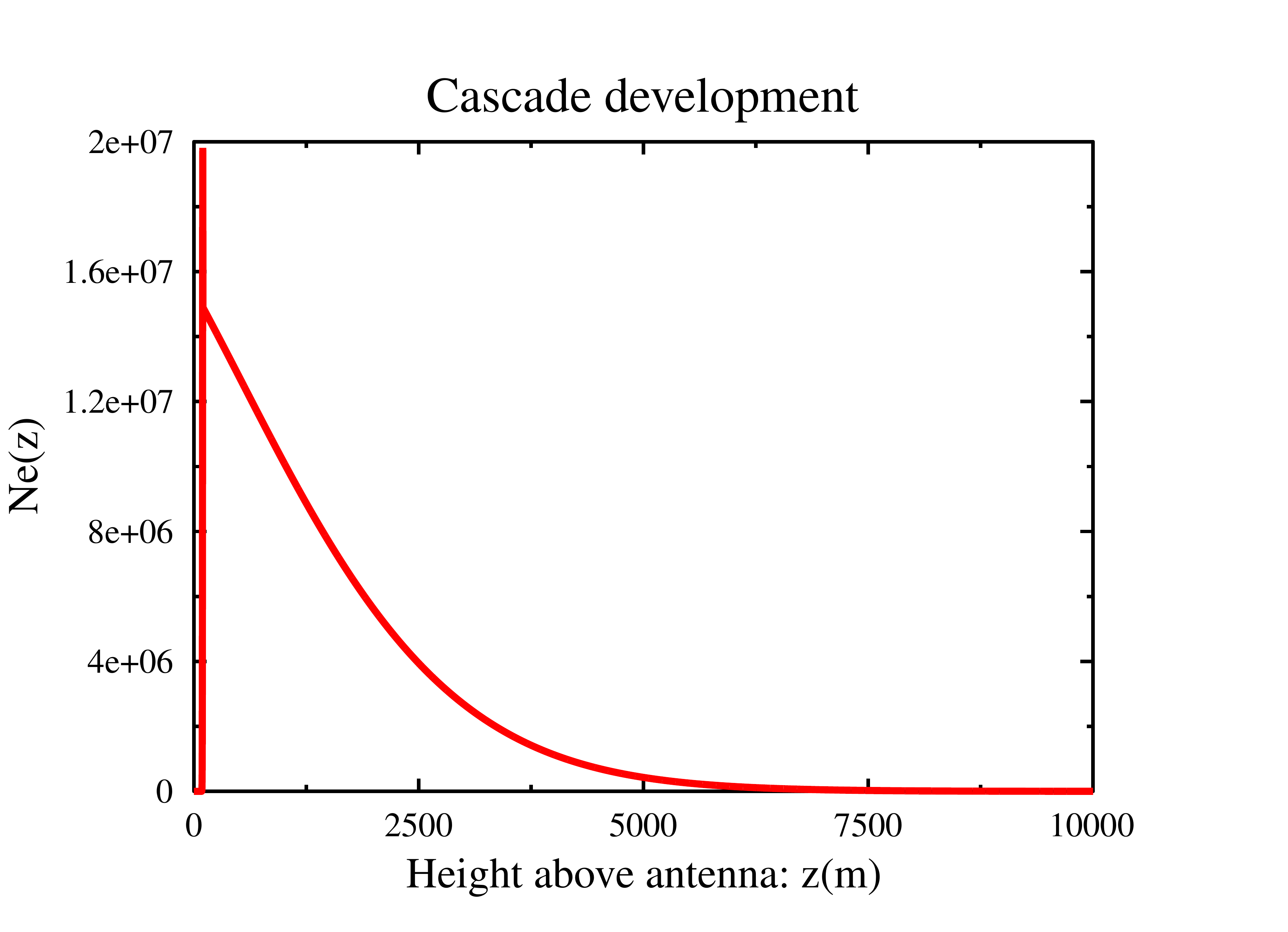}} \\
  \subfloat[]{\figlab{profileb}\includegraphics[width=.5\textwidth]{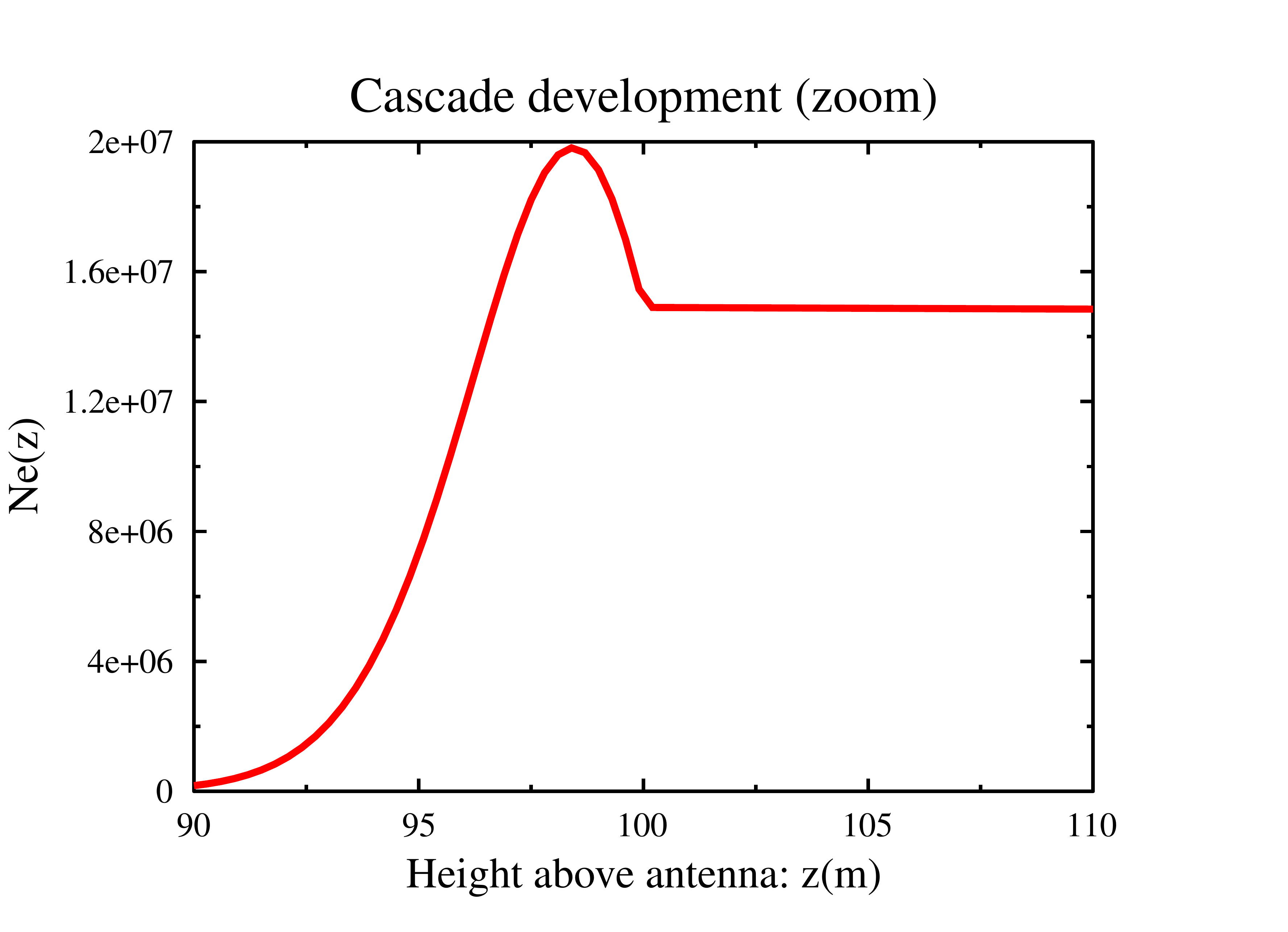}}
  \caption{The number of charges $N_e$ as a function of height $z$ above the antenna as simulated for a cosmic-ray air shower with primary particle energy of $10^{17}$~eV. The antenna is positioned at an observer level 2900~m above sea level in ice. The ice-air boundary is at 3000~m above sea level.}
  \figlab{profile}
\end{figure}

In the previous section, we obtained the electric field expressions for the transition radiation from a particle cascade traversing the boundary between two different media. We also considered the steady charge emission as well as the varying current emission. We now have all ingredients to solve for the the emission from a high-energy cosmic-ray air shower which penetrates a surface.

\subsection{The particle cascade}
As an example we consider a shower which is induced by a $10^{17}$~eV primary cosmic ray, where the shower will hit an ice surface. The shower profile can be expressed as a function of the penetration depth $X(g/cm^2)=\int \rho(z) \mathrm{d}z$, given by the line integral over the density which the shower has passed through. This allows us to naturally take into account for the air-ice boundary by simply writing the density as,
\beq
\rho(z)=\rho_{air}\,\theta(z-z_{b}) + \rho_{ice}\,\theta(z_{b}-z) \;,
\eeq
again using the step function $\theta(x)$. We will assume a density profile $\rho_{air}(z)=\rho_0 \exp[-C(z+z_0)]$ for an observer positioned at a height $z=z_0$ above sea level. Here $C=1.160\times 10^{-4}\;\mathrm{m^{-1}}$, and $\rho_0=1.168\times 10^{-3}\;\mathrm{g\;cm^{-3}}$~\cite{dVries10}. The ice density is assumed to be constant over the few meters in which the cascade will die out and taken as $\rho_{ice}=0.92\;\mathrm{g\;cm^{-3}}$.

Since the radiation length $X_0=36\;\mathrm{g/cm^2}$, as well as the critical energy $E_{crit}=80$~MeV for electrons is approximately equal in air and ice, we can now take a NKG approximation~\cite{Kam58,Grei65} given by,
\beq
N(X)=\frac{0.31\;\exp[(X/X_{0})(1-1.5\ln s)]}{\sqrt{\ln(E/E_{crit})}}
\eqlab{NKG}
\eeq
for the total number of particles as a function of depth. The shower age $s$ is given by~\cite{Hue03},
\beq
s(X)=\frac{3X/X_0}{(X/X_0)+2\ln(E/E_{crit})}\;.
\eqlab{age}
\eeq
The excess charge as function of shower depth can be approximated by $N_{ch}(X)\approx 0.23 \,N(X)$~\cite{dVries10}. The total number of excess electrons as function of depth is shown in~\figref{profile}. Taking a geometry with the observer positioned at 2900~m above sea level with the air-ice boundary at 3000~m, the boundary at $z_b=100$~m is clearly visible.

The particle distribution in the shower front is given by the weight function $w(r,h)=\delta(r)f(h)$, where $\int \mathrm{d}h\; f(h)=1$. The radial extension of the particles in the shower front is taken to be a delta function at the shower axis. To compensate for the loss of the lateral coherence scale, we use an effective width $h_1$ for the longitudinal particle distribution in the shower front. This width therefore reflects the coherence scale due to the full extension of the shower front. The longitudinal distribution of particles in the shower front is parametrized by~\cite{Hue03,Agn03},
\beq
f(h)=(4/h_1^2)\,h\,e^{-2h/h_1}\;.
\eeq
The effective width $h_1$ is chosen to be $h_1^{air}=0.5$~m following~\cite{dVries11} for the in-air development, while for the in-ice part of the cascade a width $h_1^{ice}=0.1$~m is chosen.

\subsection{The refractive index}
In~\cite{dVries11} it was shown that in determining the radio signal from cosmic-ray air showers it is crucial to take into account a realistic index of refraction. Therefore, in the following we model the index of refraction in air by the Gladstone-Dale law,
\beq
n_{air}(z)=1+0.226\mathrm{\frac{g}{cm^3}}\rho_{air}(z) \;.
\eeq
Furthermore, in~\cite{Wer08} it was shown that the bending of the emission in air can safely be neglected. The index of refraction in ice is taken as a constant equal to
\beq
n_{ice}=1.78\;.
\eeq

\subsection{Results}
Since we now have our electric field expressions, as well as the particle distributions, we can calculate the electric field at different observer positions in ice. We consider two different boundary levels at 500~m and 3000~m above sea level. The shower profile is given in~\figref{profile} for a geometry where the air-ice boundary is 3000~m above sea level. In~\figref{profilea} the full shower profile is given. It follows that the shower hits the ice surface before it reaches its maximum. This is more clear from \figref{profileb} where we zoom in on the boundary. In the ice the shower quickly reaches its maximum and dies out within 10~meters.

We consider the emission as seen by an observer positioned at several distances, $d=40$~m, $d=80$~m, and $d=240$~m, from the shower axis, 100~m below the air-ice boundary. The obtained electric fields are shown in~\figref{e-field} a, c, and e. The full red line gives the electric field generated by the in-air development of the cascade, the striped purple line gives the transition radiation, and the dotted blue line gives the emission from the in-ice development of the cascade. Next to the obtained electric fields, in~\figref{e-field} b, d, and f we also show the total number of particles as function of height. Furthermore, in these figures we show the observer time for a signal emitted from a certain height. The full green line gives the total number of particles as function of height. It should be noted that the vertical axis is shifted by 90~m for plotting purposes. The full red line gives the emission height as function of the observer time for the in-air emission, where the striped purple line gives the same quantity for the in-ice emission.

The emission observed at $d=40$~m is shown in~\figref{e-field}a. As follows from~\figref{e-field}b, the in-air emission from large heights is observed before the emission from lower heights. For the in-ice emission, this picture is completely reversed. In-ice signals emitted from large heights are delayed by the medium, while the cascade continues to propagate with the speed of light. Hence signals emitted at later times (lower heights) arrive before signals emitted early and the observer is positioned inside the Cherenkov cone for the full in-ice emission. 

For the transition radiation, it is important to notice that the electric field as given in~\eqref{etrr} can be seen as a superposition of the emission just above, and just below the boundary which interferes destructively. The emission scales with $1/|\bigD|\sim|dt_r/dt|\sim|dz/dt|$, which is reflected in~\figref{e-field}b by the slopes of the full red and striped purple lines at the boundary. At the boundary, there will be a sudden change of the particle distributions in the shower front. To take this into account in our modeling, the emission just above the boundary is evaluated using the particle distributions for the in-air shower, where the component just below the boundary is modeled using the particle distribution for the in-ice cascade. From this point of view one might also consider the transition radiation from just above the boundary as the sudden-death signal from the emission in air, where the transition radiation from just below the boundary can be seen as the sudden appearance signal for the in-ice emission.

For an observer positioned at $d=80$~m, see~\figref{e-field}c and d, a similar picture is obtained. The in-air emission is observed over a longer time-scale since we move further away from the Cherenkov cone for the emission emitted in air. For the in-ice emission, however, we shift closer to the Cherenkov angle. It follows that the in-ice emission is observed within a much shorter time-span and becomes much stronger. The transition radiation is now dominated by the emission from just below the boundary.

Finally we consider an observer positioned at $d=240$~m. From~\figref{e-field}e and f it follows that both the in-air emission as well as the in-ice emission are observed outside the Cherenkov cone. The emission is observed over a rather long time-scale, although the in-air component starts to be rather weak. One interesting feature is that the emission just above the boundary does not arrive at the same time as the emission just below the boundary. The transition radiation component just above the boundary is highly suppressed, and arrives just before $t=1000$~ns, where the component emitted just below the boundary is much stronger and arrives at a much later time around $t=1200$~ns. The time difference arises due to the fact that the signal emitted just above the air-ice boundary will first travel a significant part of its path almost horizontally before breaking into the ice under the critical angle, which in this case is equal to the in-ice Cherenkov angle. The signal emitted just below the air-ice boundary will travel its full path through the ice and hence obtain a large delay with respect to the signal emitted just above the boundary. It should be noted that this effect occurs in the situation of a perfectly flat and smooth surface. In a realistic experiment, the emission from just above the boundary however will not be able to travel almost perfectly horizontally along the surface and hence will loose coherence and become suppressed (already in the present case it is almost negligible in magnitude). The signal emitted just below the surface will not be affected and keep its coherence.
\begin{figure*}[]
  \subfloat[]{\includegraphics[width=.5\textwidth]{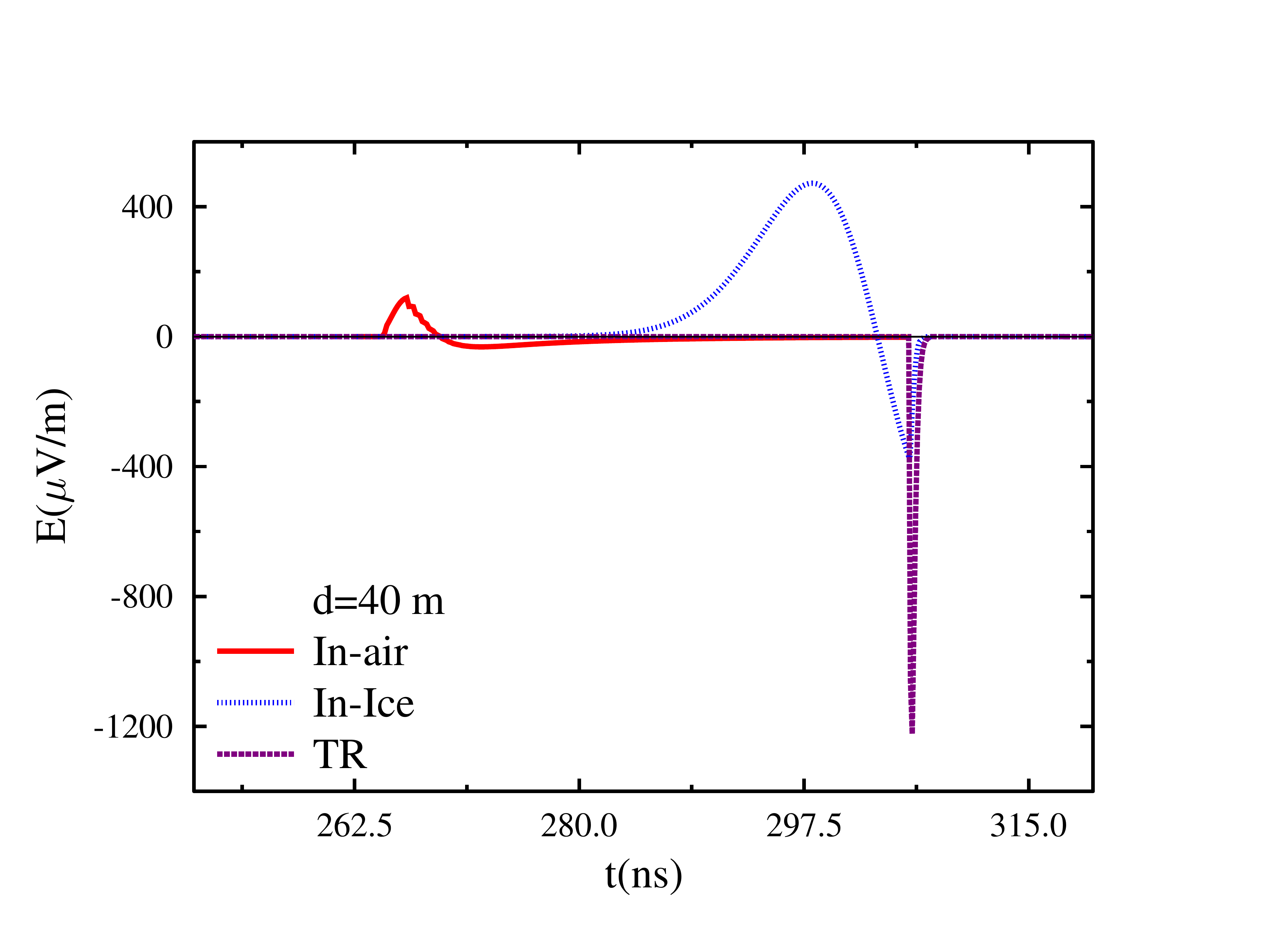}}\hfill
  \subfloat[]{\includegraphics[width=.5\textwidth]{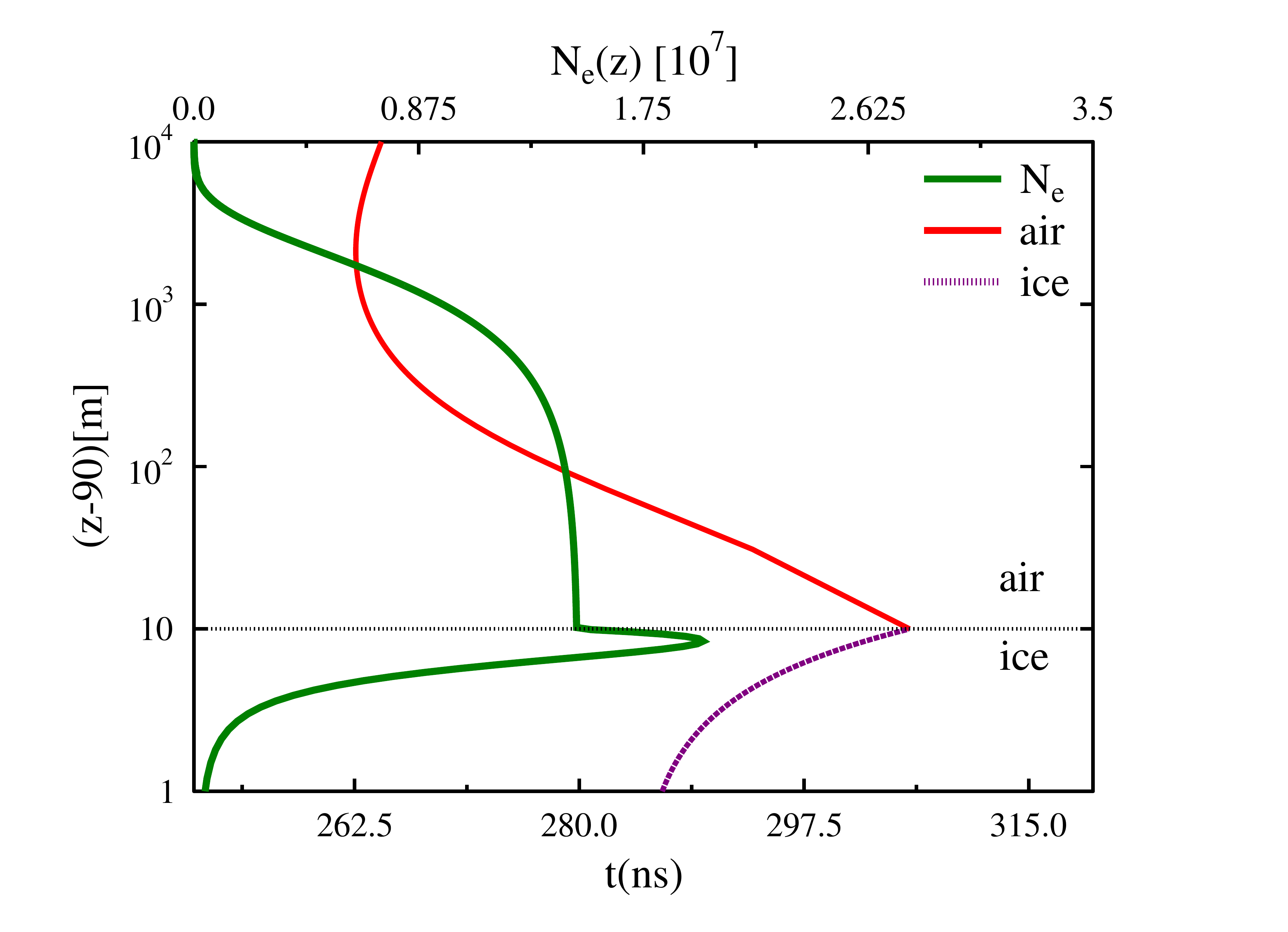}}\\
  \subfloat[]{\includegraphics[width=.5\textwidth]{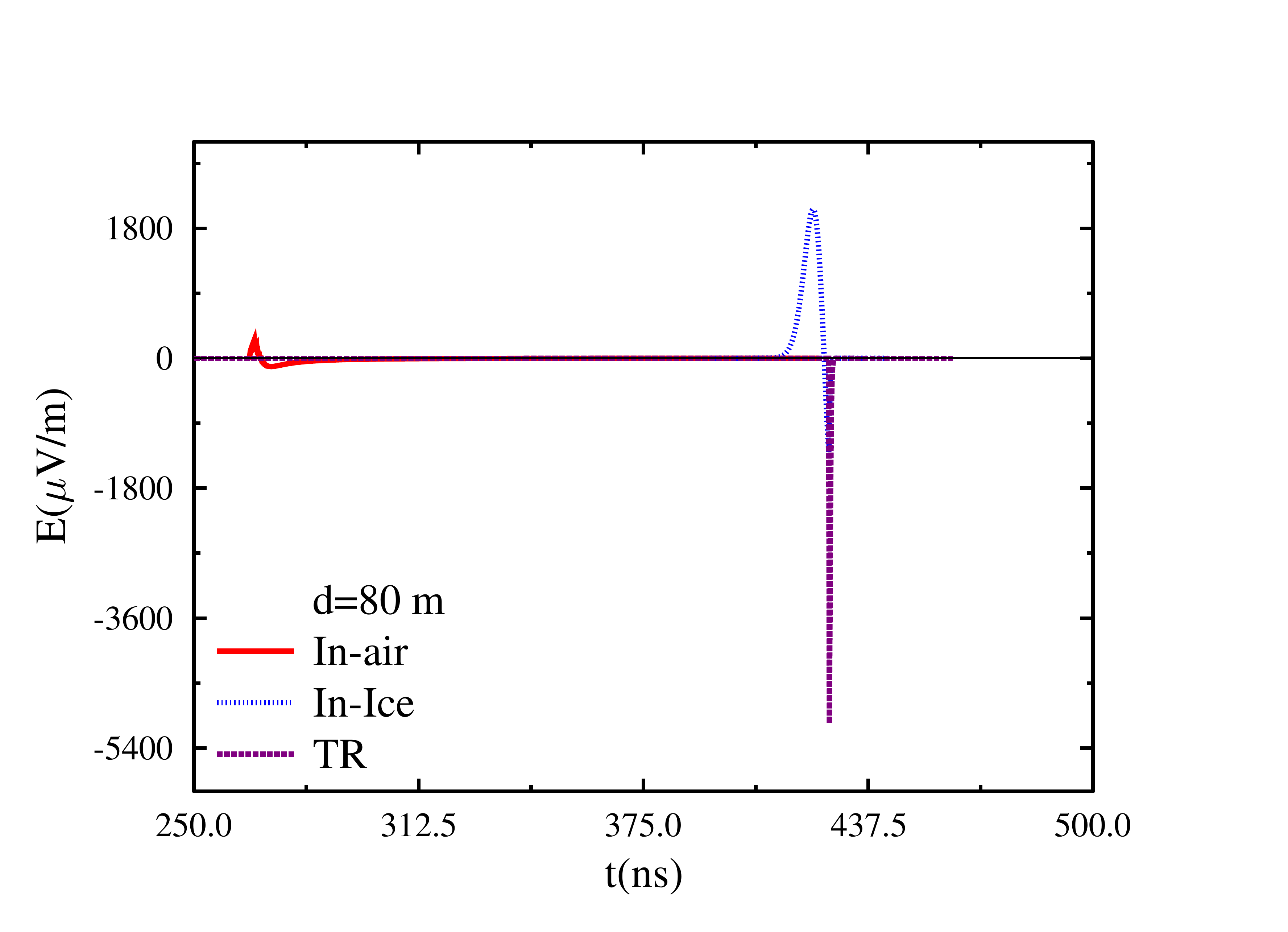}}\hfill
  \subfloat[]{\includegraphics[width=.5\textwidth]{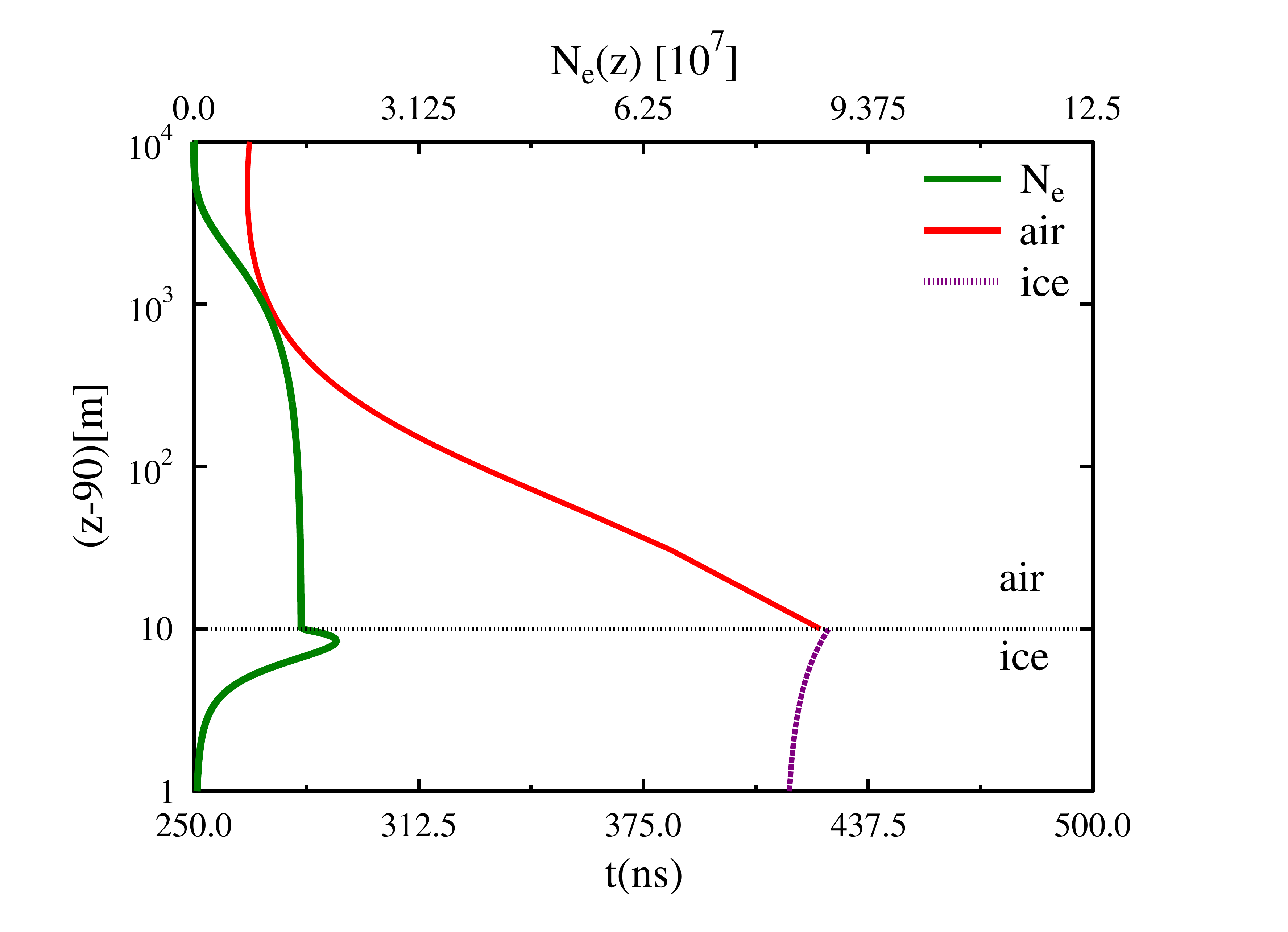}}\\
  \subfloat[]{\includegraphics[width=.5\textwidth]{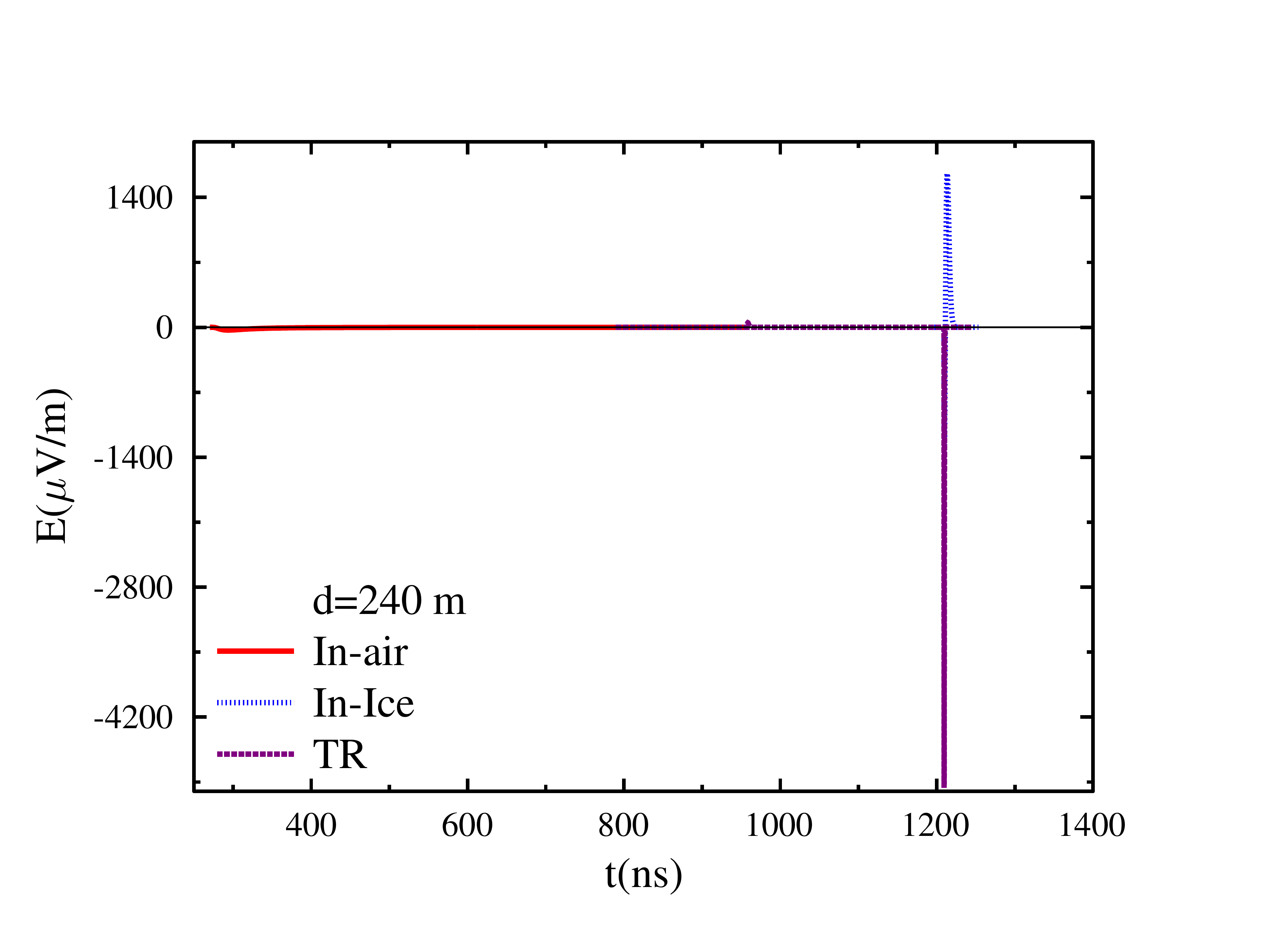}}\hfill
  \subfloat[]{\includegraphics[width=.5\textwidth]{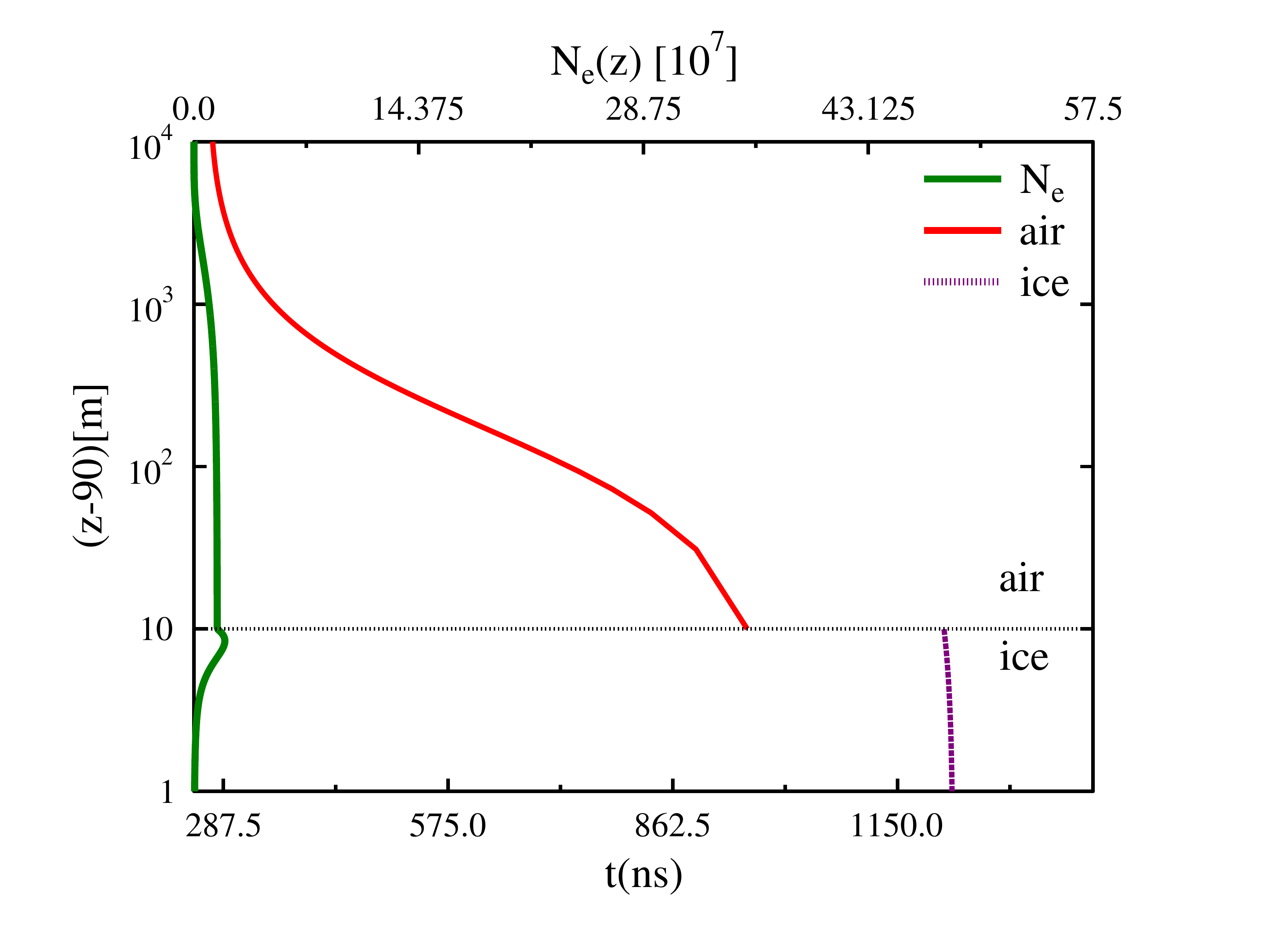}}\\
  \caption{The electric field at different observer distances equal to, a) $d=40$~m, b) $d=80$~m, c) $d=240$~m. The figures on the right show the emission height, plotted as function of the observer time. The full red line gives the emission in air, the dotted purple line gives the transition radiation, and the dashed blue line gives the in-ice emission. For the figures on the right, the total number of particles is given by the full green line.}
  \figlab{e-field}
\end{figure*}

The emission will be coherent up to relatively high frequencies. This is also seen in~\figref{e-field-freqa} where we plot the frequency spectrum of the different components of the emission when the observer is positioned at a distance of $d=240$~m. In~\figref{e-field-freqb}, we plot the frequency spectrum for the same geometry shifting the ice-air boundary to 500~m above sea level.
\begin{figure}[ht!]
  \subfloat[]{\figlab{e-field-freqa}\includegraphics[width=.5\textwidth]{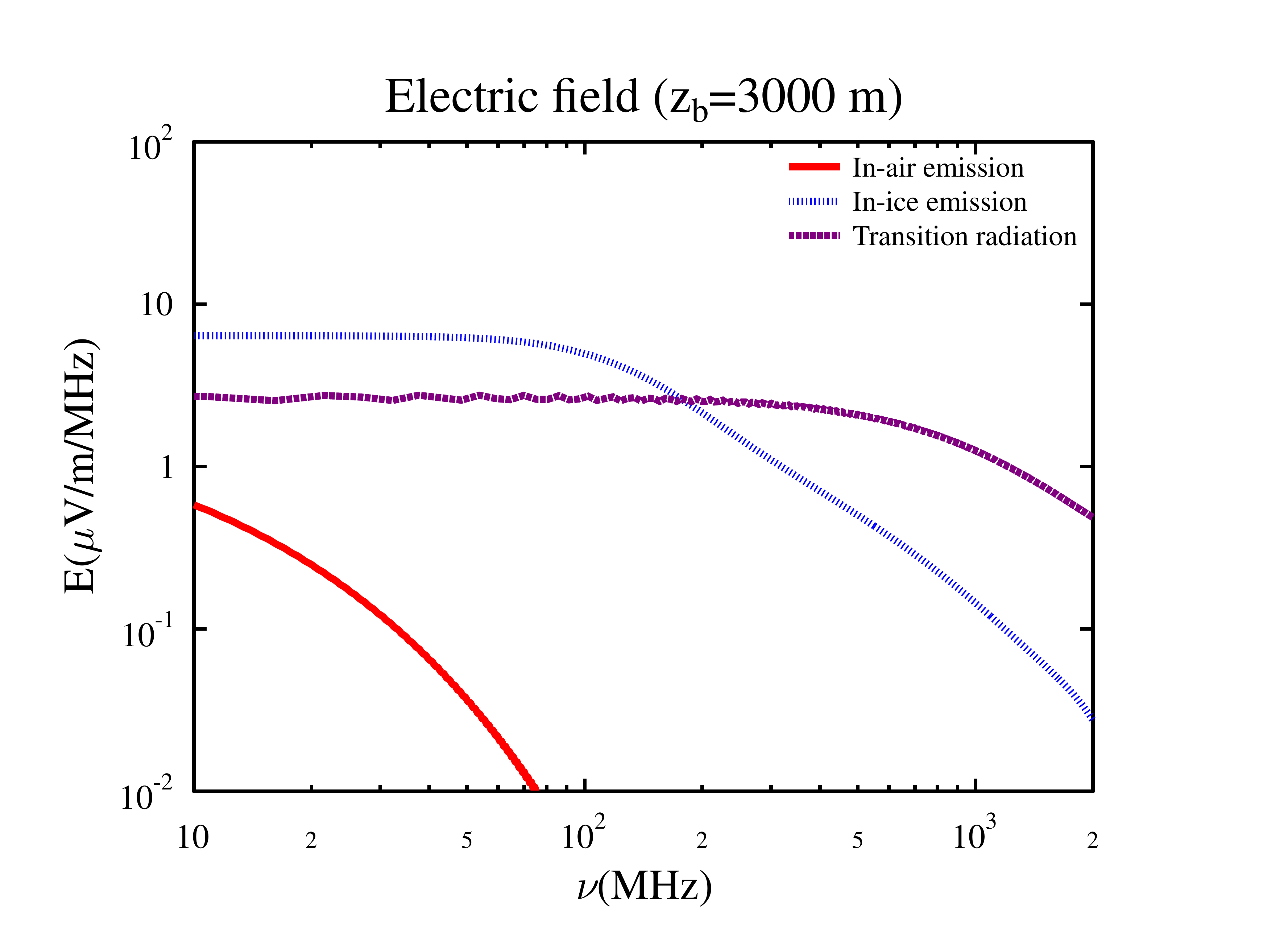}} \\
  \subfloat[]{\figlab{e-field-freqb}\includegraphics[width=.5\textwidth]{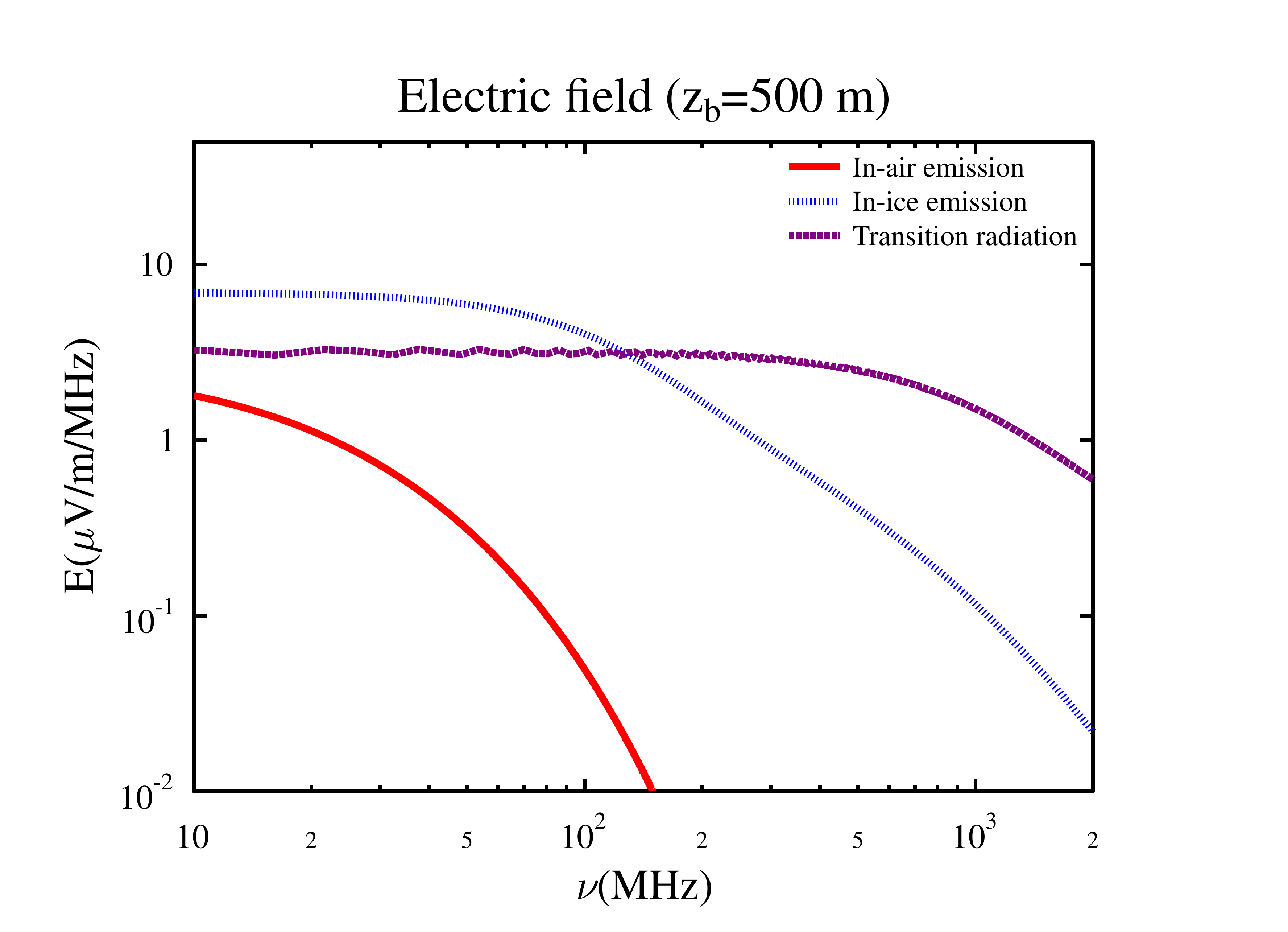}}
  \caption{The frequency spectrum of the different components to the electric field as seen by an observer positioned 100~m below the ice-air boundary and $d=250$~m from the shower axis. The simulation is performed for a $10^{17}$~eV primary energy cosmic-ray air shower.}
  \figlab{e-field-freq}
\end{figure}
Coherence of the in-ice emission as well as the in-air emission away from the Cherenkov angle is typically determined by the length of the shower trajectory leading to a suppression at the highest frequencies. The transition radiation, however, is emitted from a single point at the boundary, and hence its coherence is fully determined by the particle distributions in the shower front which gives a cut-off at relatively high-frequencies in the GHz range. Each of the several different components has a finite response at zero frequency. One should note however that the combined response of all different components vanishes at zero frequency.

In~\figref{e-field}, and~\figref{e-field-freq}, the detailed properties of the emission in time and frequency space were shown. This allows us to understand the angular distribution of the different components of the emission shown in~\figref{e-field-angle}. Here we plot the integrated absolute value of electric field $I=\int |E| dt$.
\begin{figure}[ht!]
  \subfloat[]{\figlab{e-field-anglea}\includegraphics[width=.5\textwidth]{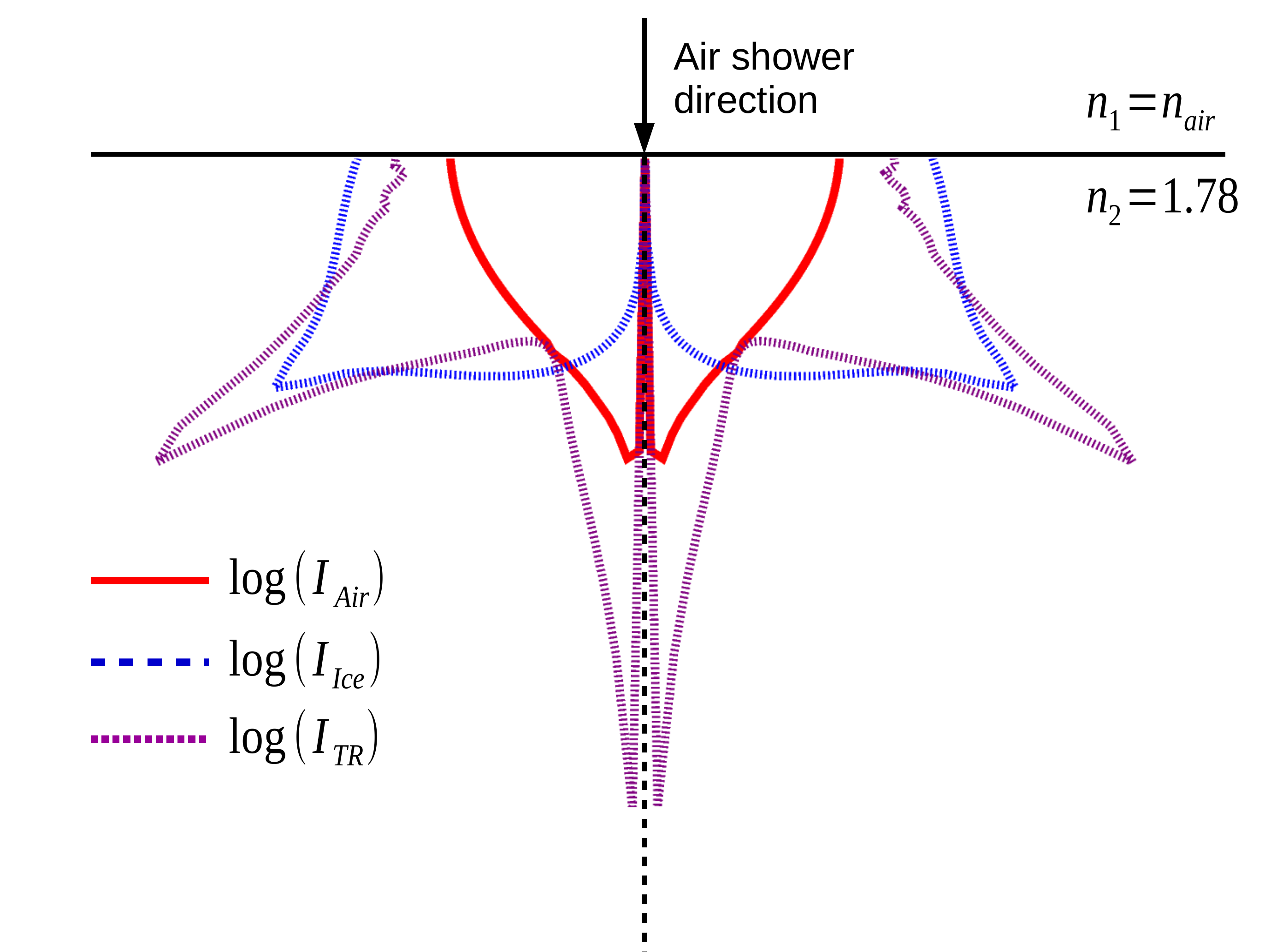}} \\
  \subfloat[]{\figlab{e-field-angleb}\includegraphics[width=.5\textwidth]{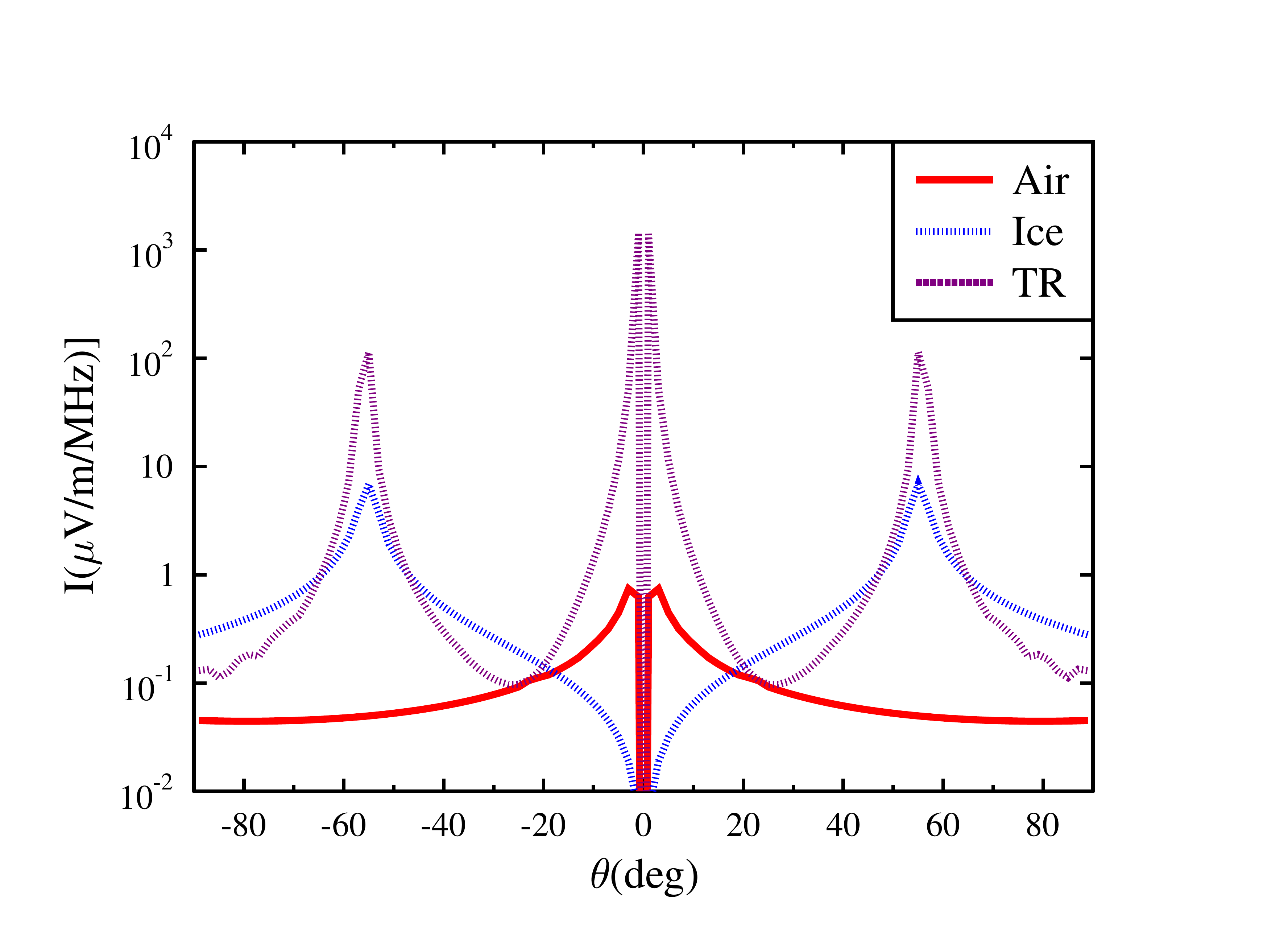}}
  \caption{The angular distribution of the integrated field for a radial observer distance of $R=\sqrt{z^2+d^2}=300$~m. The full red line gives the angular distribution for the in-air emission, the dotted purple line gives the transition radiation, and the dashed blue line shows the in-ice emission.}
\figlab{e-field-angle}
\end{figure}
It follows that the in-air emission given by the full red line in~\figref{e-field-angle}, as expected, peaks toward a highly forward angle ($\theta=0^\circ$), after which the emission drops rapidly toward larger angles. The transition radiation, shown by the dotted purple line in~\figref{e-field-angle}, shows a similar behavior as observed in~\figref{tr_angle}. There is a strong forward peak, after which the transition radiation gets suppressed due to the destructive interference between the emission just above the air-ice boundary and the emission just below the air-ice boundary. When the destructive interference is maximal, contrary to~\figref{tr_angle}, the field does not vanish completely. This is due to the different particle distributions for the in-air emission and the in-ice emission which is taken into account for in the evaluation of the transition radiation. The in-ice emission, as expected, peaks at the in-ice Cherenkov angle. Nevertheless, the emission pattern is rather broad toward smaller angles due to the longitudinal extent of the cascade.

\subsection{Zenith angle dependence}
Up to now we only considered a perpendicular incoming cosmic-ray induced air shower. Since a shower coming in under a finite zenith angle can be treated by a direct rotation of the geometry, we do not expect the emission to change significantly. One effect that is to be expected for an inclined shower, or in case of a non-perpendicular boundary, is that the transition radiation from different radial parts of the shower is emitted at different times. In case of an observer positioned underneath the shower axis this will lead to a small additional spread in the arrival time of signals emitted from different positions in the shower front, and hence a slight decrease of pulse-strength can be expected. For an observer positioned away from the shower axis however, this effect is reversed, leading to a slightly enhanced pulse-strength in the detector. Furthermore, it should also be noted that for more inclined geometries, a larger part of the signal created in air will be reflected off of the surface suppressing the in-air emission over the in-ice emission even more.

The most important effect, however, will be due to the change of the total number of charges hitting the air-ice boundary. Since for larger zenith angles the shower traverses a longer distance through air, the total number of particles actually hitting the air-ice boundary changes. Other effects influencing the total number of particles hitting the air-ice surface are the chemical composition and the energy of the primary cosmic ray. An iron induced shower typically develops earlier in the atmosphere than a proton induced shower, where cosmic rays of higher energy typically peak deeper in the atmosphere. In~\figref{N_zenith} we show the total number of particles hitting the air-ice surface for boundary layers at $z_b=3000$~m (full lines) and $z_b=500$~m (dotted lines) for a typical proton shower with a primary energy of $E_p=10^{17}$~eV (red lines) and $E_p=10^{18}$~eV (blue lines). 

For an air-ice boundary at $z_b=3000$~m, the air shower is still below shower maximum for both considered energies. It follows that the total number of particles peaks at a zenith angle of approximately $\theta\approx 40-50$~degrees, where the air shower is fully developed at the boundary. For larger zenith angles, the total number of particles hitting the air-ice boundary becomes smaller, and for zenith angles larger than $\theta \gtrsim 60$ degrees the shower dies out before hitting the air-ice boundary. Hence no transition radiation and in-ice emission will be observed for showers at zenith angles larger than approximately $\theta\approx60$~degrees. 
\begin{figure}[!ht]
\centerline{
\includegraphics[width=.5\textwidth, keepaspectratio]{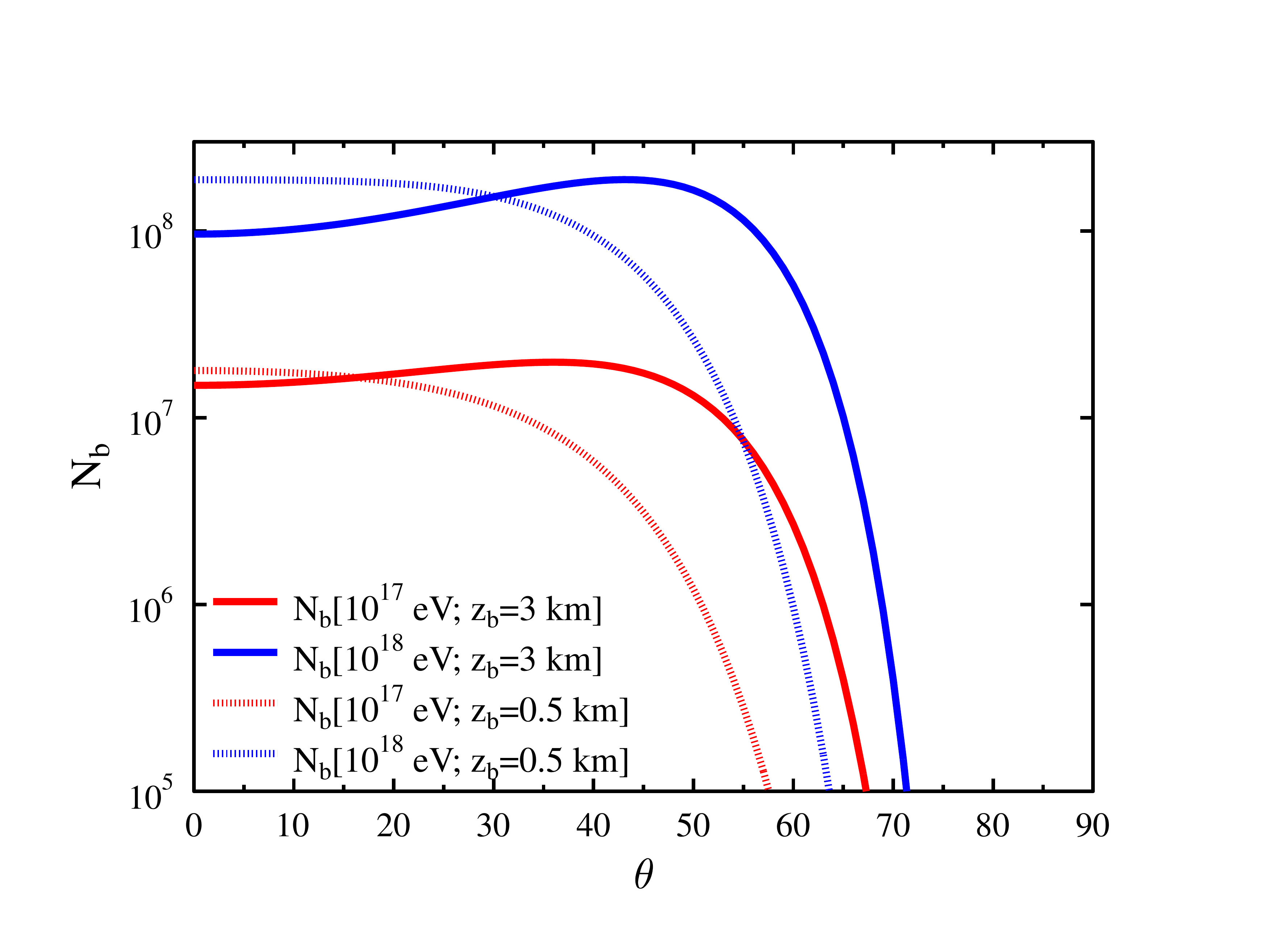}}
\caption{The total number of particles hitting the air-ice surface for boundary layers at $z_b=3000$~m (full lines) and $z_b=500$~m (dotted lines) for a typical proton shower with a primary energy of $E_p=10^{17}$~eV (red lines) and $E_p=10^{18}$~eV (blue lines). }
\figlab{N_zenith}
\end{figure}

\subsection{Cosmic-ray air shower or neutrino induced cascade?}
One important question to consider is how the cosmic-ray air shower signal compares to the emission from a neutrino induced cascade in ice. This is shown in~\figref{e-field-comp}. For the cosmic-ray air shower signal, we consider both the in-ice emission as well as the transition radiation component just below 
the boundary. As follows from~\figref{e-field}, the in-air emission is very small and will therefore be ignored for this comparison.

\begin{figure}[!ht]
\centerline{
\includegraphics[width=.5\textwidth, keepaspectratio]{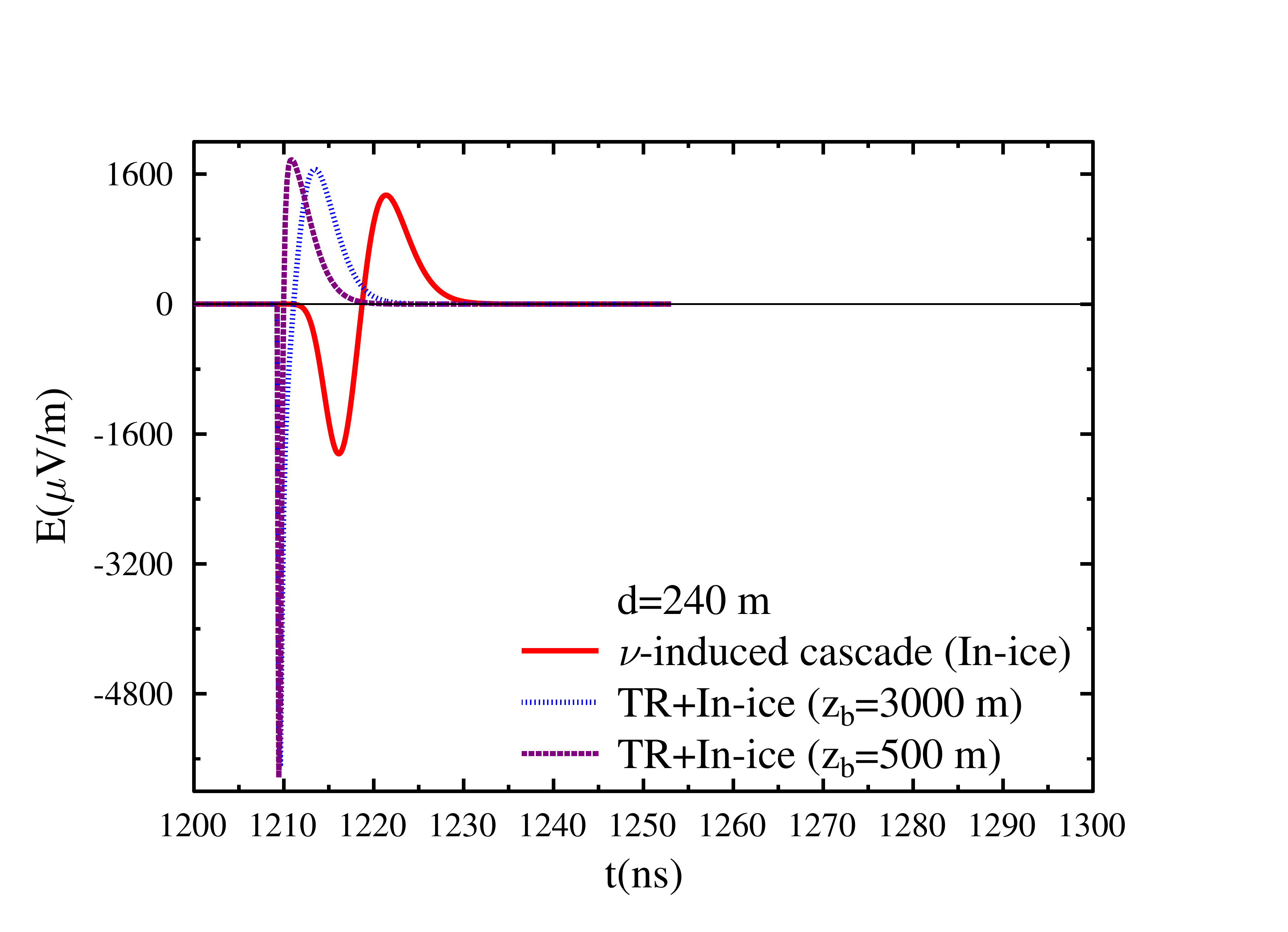}}
\caption{The electric field from a neutrino induced particle cascade in ice (full red line), compared to the transition radiation just below the boundary layer and the in-ice emission from a vertical, proton induced, cosmic-ray air shower of the same energy. The electric field is shown for ice-air boundaries equal to $z_{b}=3000$~m (dashed blue line) and $z_b=500$~m (dotted purple line). The observer is positioned at a lateral distance $d=250$~m perpendicular to the shower axis.}
\figlab{e-field-comp}
\end{figure}

The neutrino induced cascade is modeled by taking a $10^{17}$~eV primary neutrino interacting at the surface of the air-ice boundary, after which the cascade develops in ice. The observer is positioned 100~m below the ice surface at a distance $d=240$~m perpendicular to the shower axis. The effective width of the particle distribution is taken as $h_1=0.1$~m, approximately corresponding to the dimensions of the cascade front in ice.

It follows that the emission from a typical proton induced high-energy cosmic ray air shower hitting the ice surface is of similar magnitude compared to the Askaryan signal from a neutrino induced particle cascade in ice of the same energy. Since the total number of particles hitting the air-ice interface is slightly larger at $z_b=500$~m for the considered geometry of a vertical cosmic-ray air shower (see~\figref{N_zenith}), the transition radiation gets slightly enhanced with respect to the emission seen at $z_b=3000$~m.

The direct consequence is that the radio emission from a high-energy cosmic ray air shower will be very hard to distinguish from the emission of a high-energy neutrino induced particle cascade without directional information about the signal. Another possible method to separate the cosmic-ray air shower from the neutrino induced cascade might be obtained by adding a surface veto to the array.

\section{Summary and conclusions}
We derived an analytical expression for coherent transition radiation from a particle bunch with a net charge which is traversing from one medium to another. In addition to the transition radiation also the emission due to a highly-relativistic steady charge and a time-varying current are given.

As a first application we calculate the radio emission from a cosmic-ray-induced air shower hitting an ice layer before the shower has died out. It is shown that a relatively strong transition radiation component can be expected from an air shower when taking an air-ice boundary in the range between 500-3000~m above sea level. The emission from such a cosmic-ray air shower is calculated to be of similar strength as the Askaryan signal obtained from an in-ice cascade induced by a GZK-neutrino of similar energy. Furthermore, the polarization of the transition radiation will be similar to the polarization of the Askaryan signal. It follows that without directional information or a surface veto, it will be very hard to distinguish between both signals. Therefore, the emission from cosmic-ray air showers might induce a significant (background) signal in the currently operating Askaryan neutrino detectors at Antarctica.

Next to the discussed application for cosmic-ray air showers hitting a dense medium, the transition radiation from neutrino induced particle cascades traveling from a dense medium to air or vacuum is also expected to give a strong signal. This signal might be a promising probe to detect high-energy neutrino-induced particle cascades escaping dense media. A more detailed calculation for this component will be given in future work.

\section{Acknowledgments}
The authors wish to thank The Flemish Foundation for Scientific Research (FWO-12L3715N - Krijn D. de Vries), and the FWO Odysseus program (G.0917.09. - N. van Eijndhoven), and the FRS-FNRS (Aongus \'O Murchadha) for making this research possible.


\begin{thebibliography}{00}



%
\bibitem{ANITA} P.W. Gorham~et al., Phys.~Rev.~Lett. \VYP{103}{2009}{051103}
\bibitem{ARA} P. Allison~et al., ARA~Collaboration, Astropart.~Phys.~\VYP{35}{2012}{457-477}
\bibitem{ARIANNA} ARIANNA Collaboration,  Proc.~32nd~ICRC~Rio~De~Janeiro, Brasil, to~be~published
\bibitem{Ask62} G.A. Askaryan, Sov.~Phys.~JETP~\VYP{14}{1962}{441};~\VYP{21}{1965}{658}
\bibitem{Zas92}  E. Zas, F. Halzen, and T. Stanev, Phys.~Rev.~D~\VYP{45}{1992}{362}
\bibitem{Mun97}  J. Alvarez-Mu\~niz and E. Zas, Phys.~Lett.~B~\VYP{411}{1997}{218}
\bibitem{Sal01} D. Saltzberg~et al., Phys.~Rev.~Lett.~\VYP{86}{2001}{2802}

\bibitem{Mar11} V. Marin, CODALEMA~Collaboration, Proc.~32nd~ICRC, Beijing, China
\bibitem{Aab14} A. Aab~et al., Pierre Auger Collaboration, Phys.~Rev.~D\VYP{89}{2014}{052002}
\bibitem{Sch15} P. Schellart~et al., arXiv:1406.1355

\bibitem{Kah66} F.D. Kahn and I. Lerche, Proc. Royal Soc. London~\VYP{A289}{1966}{206}
\bibitem{Sch08} O. Scholten, K. Werner, F. Rusydi, Astropart.~Phys.~\VYP{29}{2008}{94-103}
\bibitem{Fal05} H. Falcke~et al., Nature~\VYP{435}{2005}{313}
\bibitem{Ard06} D. Ardouin~et al., Astropart.~Phys.\VYP{26}{2006}{341}

\bibitem{Nel15} A. Nelles~et al., Astropart.~Phys.~\VYP{65}{2015}{11-21}
\bibitem{Bui15} S. Buitink~et al., Phys.~Rev.~D~\VYP{90}{2014}{082003}

\bibitem{Alv12} J. Alvarez-Muniz, W. R. Carvalho, Jr., E. Zas, Astropart.~Phys.~\VYP{35}{2012}{325-341}
\bibitem{Mar12} V. Marin, B. Revenu, Astropart.~Phys.~\VYP{35}{2012}{733-741}
\bibitem{Hue12} T. Huege, M. Ludwig, C. James, AIP~Conf.~Proc.~\VYP{1535}{2013}{128-132}
\bibitem{Wer12} K. Werner, K.D. de Vries, O. Scholten,~Astropart.~Phys.~\VYP{37}{2012}{5-16}


\bibitem{Numoon} O. Scholten~et al., Phys.~Rev.~Lett.~\VYP{103}{2009}{191301}
\bibitem{Jam09} C.W. James et al., MNRAS~\VYP{410(2)}{2011}{885-889}
\bibitem{Veen10} S. ter Veen~et al., Phys.~Rev.~D~\VYP{82}{2010}{103014}






\bibitem{Grei66} K. Greisen, Phys.~Rev.~Lett.~\VYP{16}{1966}{748}
\bibitem{Zat66} G.T. Zatsepin, V.A. Kuzmin, Pis’ma~Zh.~Eksp.~Teor.~Fiz.~\VYP{4}{1966}{114}

\bibitem{Dag89}  R. Dagesamanski and I. Zheleznyk, Sov.~Phys.~J.E.T.P.\VYP{50}{1989}{233}
\bibitem{Han96} T. Hankins, R. Ekers, and J. O’Sullivan, MNRAS~\VYP{283}{1996}{1027}
\bibitem{FORTE} H. Lethinen~et al., Phys.~Rev.~D~\VYP{69}{2004}{013008}
\bibitem{Gor04} P. Gorham~et al., Phys.~Rev.~Lett.~\VYP{93}{2004}{41101}
\bibitem{Ber05} A. Beresnyak~et al., Astronomy~Reports~\VYP{49}{2005}{127}



\bibitem{I3_2013sc} IceCube~Collaboration, Science~\VYP{342}{2013}{1242856}


\bibitem{Mar86} M.A. Markov, I.M. Zheleznykh, Nucl. Instrum. Meth. Res.~A~\VYP{248}{1986}{242}
\bibitem{Mar14} B. Revenu, V. Marin, arXiv:1211.3305


\bibitem{dVries10} K.D. de Vries, A.M. van den Berg, Olaf Scholten, Klaus Werner, Astropart.~Phys.~\VYP{34}{2010}{267}
\bibitem{Jackson} J.D. Jackson, Classical Electrodynamics, Wiley, New York, 1999

\bibitem{Wer08} K. Werner, O. Scholten,~Astropart.Phys.~\VYP{29}{2008}{393}



\bibitem{Gin90} V.L. Ginzburg, V.N. Tsytovich, Transition Radiation and Transition Scattering, Adam Hilger Press, New York, 1990
\bibitem{Jam11} C.W. James, H. Falcke, T. Huege, M. Ludwig, Phys.~Rev.~E~\VYP{84}{2011}{056602}
\bibitem{Shi94} Y. Shibata~et. al, Phys.~Rev.~E~\VYP{50}{1994}{1479}
\bibitem{Gor20} P.W. Gorham~et. al, Phys.~Rev.~E~\VYP{62}{2000}{8590}

\bibitem{Kam58} K. Kamata, J. Nishimura, Suppl.~Progr.~Theoret.~Phys.~\VYP{6}{1958}{93}
\bibitem{Grei65} K. Greisen, in: J.G. Wilson (Ed.), Prog.~Cosmic~Ray~Phys., vol. III, North~Holland, Amsterdam, 1965, p. 1


\bibitem{Hue03} T. Huege, H. Falcke, Astronomy~\&~Atrophys.~\VYP{19}{2003}{412}
\bibitem{Agn03} G. Agnetta~et al., Astropart.~Phys.\VYP{6}{2003}{301}
\bibitem{dVries11}K.D. de Vries, A.M. van den Berg, O. Scholten, K. Werner, Phys.~Rev.~Lett.~\VYP{107}{2011}{061101}





















\end{thebibliography}
\end{document}